\documentclass[12pt]{iopart}
\usepackage{iopams}  
\usepackage{natbib}
\usepackage{har2nat}
\usepackage{graphicx}		

\usepackage[usenames, dvipsnames]{color}

\begin{document}

\title[Generating Buoyant Loops]{Generating buoyant magnetic flux ropes in solar-like convective dynamos}

\author{Nelson, N.~J.$^1$ and Miesch, M.~S.$^2$}

\address{$^1$ Los Alamos National Laboratory, P.O. Box 1663, T-2, MS-B283, Los Alamos, NM 87545}
\address{$^2$ High Altitude Observatory, National Center for Atmospheric Research, 3080 Center Green Dr., Boulder, CO 80301}

\ead{njnelson@lanl.gov}

\begin{abstract}
Our Sun exhibits strong convective dynamo action which results in magnetic flux bundles emerging through the stellar surface as magnetic spots. Global-scale dynamo action is believed to generate large-scale magnetic structures in the deep solar interior through the interplay of convection, rotation, and shear. Portions of these large-scale magnetic structures are then believed to rise through the convective layer, forming magnetic loops which then pierce the photosphere as sunspot pairs. Previous global simulations of 3D MHD convection in rotating spherical shells have demonstrated mechanisms whereby large-scale magnetic wreaths can be generated in the bulk of the convection zone. Our recent simulations have achieved sufficiently high levels of turbulence to permit portions of these wreaths to become magnetically buoyant and rise through the simulated convective layer through a combination of magnetic buoyancy and advection by convective giant cells. These buoyant magnetic loops are created in the bulk of the convective layer as strong Lorentz force feedback in the cores of the magnetic wreaths dampen small-scale convective motions, permitting the amplification of local magnetic energies to over 100 times the local kinetic energy. While the magnetic wreaths are largely generated the shearing of axisymmetric poloidal magnetic fields by axisymmetric rotational shear (the $\Omega$-effect), the loops are amplified to their peak field strengths before beginning to rise by non-axisymmetric processes. This further extends and enhances a new paradigm for the generation of emergent magnetic flux bundles, which we term turbulence-enabled magnetic buoyancy.
\end{abstract}

\submitto{Plasma Physics and Controlled Fusion }

\section{Magnetic Spots on Sun-like Stars}

Dynamo action in sun-like stars provides an important physical laboratory for understanding magnetic self-organization in highly turbulent systems. Solar and stellar magnetic activity is dominated by the emergence and evolution of highly concentrated flux ropes which emerge through the photosphere and into the atmosphere. Sunspots exhibit ordered collective behaviors in the form of the solar magnetic activity cycle with its well-established observational features such as the 11-year polarity reversal timescale, the migration of active latitudes, Hale's polarity law, and numerous others \citep{Hathaway2010}. These collective behaviors point to the global solar dynamo as the source of the magnetic loops which become sunspot pairs. The emergence and ejection of magnetic flux also plays a key role in many dynamo models either by creating poloidal magnetic field via the emergence and dispersal of magnetic flux (known as the Babcock-Leighton mechanism) or by regulating the magnetic helicity budget \citep{Blackman2003}. Similar cycles of magnetic activity and collective starspot behaviors are beginning to be observed on a wide variety of sun-like stars \cite[e.g.,][]{Berdyugina2005, Huber2012, Llama2012, Metcalfe2013}.

Numerical models have also made significant progress in examining components of the dynamo and flux emergence processes that may operate in sun-like stars. The fundamental challenge to these models is the vast range of scale employed by processes of interest in stellar magnetism. This has led to three philosophical approaches. First, two-dimensional (2D) mean-field models seek to reduce the problem to the study axisymmetric fields and global behaviors. This demands the parameterization of essential three-dimensional (3D) effects. Mean-field models have been used extensively to examine the modes of large-scale dynamo action in the Sun \citep{Charbonneau2010}. Many mean-field models which can achieve realistic solar cycles and are used to predict the amplitude and timing of future solar cycles rely on the Babcock-Leighton mechanism \citep[e.g.,][]{Dikpati2006, Choudhuri2007}. The Babcock-Leighton mechanism is a heavily parameterized version of buoyant magnetic transport which transforms toroidal magnetic fields generated through shear amplification near the base of the solar convection zone into poloidal fields near the solar surface. Attempts to investigate this mechanism in more realistic settings have met with more limited success \citep{Durney1995, Miesch2012}.

The second philosophical approach to modeling elements of convective dynamo action and flux emergence uses local 3D models which seek to capture moderate or small scale behaviors that may be inaccessible to global scale models. In a ``bottom-up'' approach, local models often use Cartesian geometries and can be tailored to include only the needed physics for a given scale and location in the Sun. Local models have been used with great success to investigate the formation of buoyant magnetic structures using forced shear layers \citep{Cline2003, Vasil2009, Guerrero2011} and the structure and evolution of photospheric active regions \citep{Cheung2010, Rempel2011}. 

Local cartesian simulations have also demonstrated the spontaneous
generation of coherent flux structures in turbulent MHD flows by
mechanisms other than differential rotation.  Of particular note is
the negative effective magnetic pressure instability (NEMPI) first
proposed by \citet{Kleeorin1989} and recently demonstrated in
simulations of forced turbulence by \citet{Kemel2013}.  Here
coherent flux structures are generated through the suppression of
turbulent pressure by the magnetic field and the resulting compression
and confinement of the coherent field.  Though similar processes may
be occurring in our simulations, the coherent flux structures we
describe here exhibit a density deficit relative to their surroundings
rather than a density enhancement as expected from NEMPI.  Furthermore, this density deficit produces a buoyant acceleration which, together with
advection, contributes to the rise of the loops through the convection
zone.  As we stress below, both rotational shear and turbulence play
a role in the spontaneous formation of these structures, making them
in some sense a hybrid of those seen, for example, by
\citet{Guerrero2011} and \citet{Kemel2013}.

The presence of a photosphere and rarified corona can also promote the
spontaneous generation of coherent magnetic loops with a vertical
orientation at the surface.  Positive feedback between radiative
cooling and MHD induction can promote the coalescence of vertical
fields \citep{Cheung2008, Stein2012} but it is not
necessary; \citet{Warnecke2013} have demonstrated that even
adibatic simulations of forced turbulence can form bipolar magnetic
structures in the presence of a coronal envelope.  These results
may have bearing on the ultimate fate of the rising loops we
describe here.  In particular, if magnetic flux on the Sun emerges
as relatively diffuse loops such as the structures described here,
then photospheric processes may help them coalesce into more
concentrated bipolar active regions after emergence.

The third philosophical approach seeks to employ global 3D models in a ``top-down'' manner. Global models use spherical geometry and realistic stratification in an effort to achieve correct large-scale behavior and then include as broad a range of scales as computationally feasible. Global models require special attention to be devoted to the impact of unresolved scales. This can be done through either explicit or implicit Large-Eddy Simulation (LES) frameworks \citep[see][]{Grinstein2007}. convective dynamo models have received strong recent attention as four different codes have yielded models with global-scale toroidal magnetic structures, cycles of magnetic activity, and reversals in global magnetic polarity \citep{Browning2006, Ghizaru2010, Gastine2012, Kapyla2012a}. A related approach uses 3D magnetohydrodynamic (MHD) or thin flux-tube models where magnetic structures are imposed and then allowed to emerge through the convective layer \citep[e.g.,][]{Fan2009, Weber2012a, Jouve2013}.

Challenged and inspired by both theoretical and observational advances in understanding magnetic activity on sun-like stars, we have undertaken a series of 3D MHD models of global convective dynamo action in the deep interiors of sun-like stars. These models have shown that convective dynamos can create global-scale toroidal magnetic wreaths in the bulk of the convection zone \citep{Brown2010} and that these wreaths can lead to reversals in magnetic polarity and cycles of magnetic activity as rotation is increased \citep{Brown2011} or the level of turbulence is increased \citep{Nelson2013a}. For our least diffusive simulations, portions of these wreaths have become buoyant and risen coherently through our simulated domain, forming buoyant magnetic loops \cite{Nelson2011}. Collectively these loops exhibit statistical behaviors that mimic those observed in sunspots \citep{Nelson2014}.

In studying the buoyant magnetic loops created by our simulations, we have previously focused on their properties at maximum radial extent, the dynamics of their ascents, and the relation between the appearance of loops and the global-scale axisymmetric fields. Here we address the important question of how buoyant magnetic flux ropes are formed. In many ways our simulations are uniquely suited to address this question as previous models have used a forced shear-layer to generate their buoyant magnetic structures or have inserted {\it ad hoc} magnetic structures by hand. Here we directly test the assumptions common to many previous models that buoyant magnetic fields are generated through axisymmetric toroidal shear acting on poloidal magnetic fields and that dissipation of these structures is primarily caused by small-scale turbulent advective mixing \citep{Fan2009, Guerrero2011, Jouve2013}. Our analysis shows that for a small number of sample buoyant magnetic loops, the creation mechanism is not primarily driven by the axisymmetric differential rotation profile but rather depends heavily on local toroidal shear amplification. We also confirm that the dissipation mechanism is primarily due to small-scale turbulent advection.

\section{Simulation Overview}

Here we report on the results of 3D MHD simulations of turbulent convective dynamo action in a rotating spherical shell spanning the bulk of the convection zone of a sun-like star. These simulations have been conducted using the ASH code \citep{Clune1999, Brun2004}. ASH uses 1D stellar models to create a radial profile which is used as the reference state and is kept fixed throughout the temporal evolution of the simulation. Convective motions and magnetic fields are treated as perturbations around the background state. ASH uses spherical harmonics in the horizontal directions and Chebyshev polynomials in the radius to achieve spectral accuracy over long time evolutions. Our domain extends from the base of the convection zone at $0.72 R_\odot$ to near the photosphere at $0.97 R_\odot$ with a density contrast of 25 between the upper and lower boundaries. This simulation has a dimensional bulk rotation rate of 1240 nHz (thus a period of 9.3 days), or three times the current solar rate. In spite of the faster rotation rate, the results presented here may be largely revenant to solar behavior as the relevant non-dimensional parameter which gauges the relative level of rotational influence, the Rossby number, is small both here and in the bulk of the solar interior. 

\begin{figure}[t]
\begin{center}
  \includegraphics[width=0.5\linewidth]{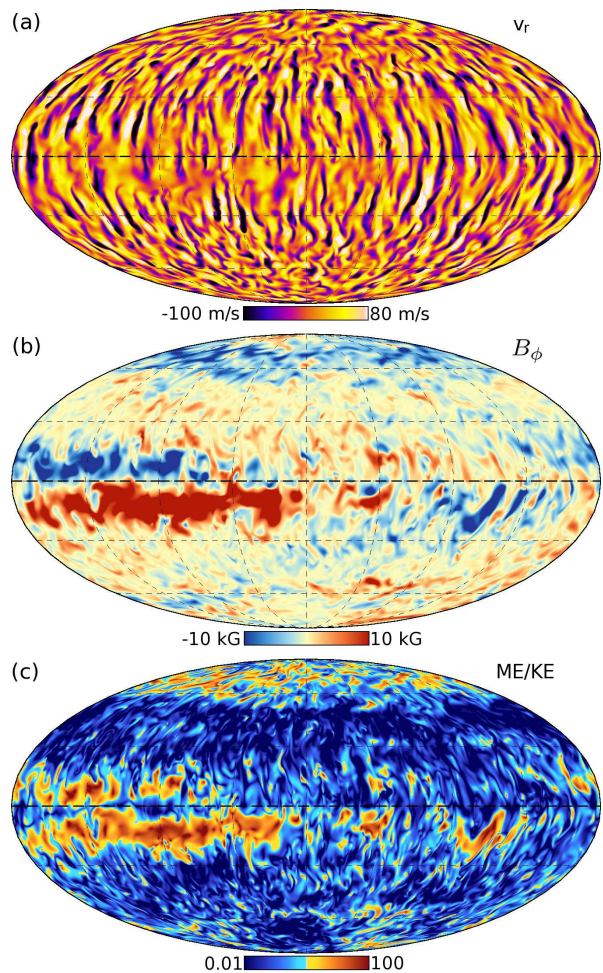}
  \caption{(a) Snapshot of radial velocities $v_r$ (dark downflows, light upflows) at $0.79 R_\odot$ for case S3 on a spherical surface shown in Mollweide projection. Convection is rotationally aligned at low latitudes and isotropic at high latitudes. (b)~Companion snapshot of longitudinal magnetic field $B_\phi$ at the same depth and instant in time. Two strong wreath segments are present with a negative-polarity (blue) wreath dominating the northern hemisphere and a positive-polarity (red) wreath dominating the southern hemisphere. (c)~Companion snapshot of the local ratio of magnetic to kinetic energies shown on a logarithmic scale.
  \label{fig:Shells}}
  \end{center}
\end{figure}

A number of simulations in this parameter regime have been conducted to explore the effects of viscous, thermal, and resistive diffusion on our solutions \citep{Nelson2013a}. Here we will focus on one simulation, case S3. Case S3 uses a fixed grid with 192 points in radius, 512 in latitude, and 1024 in longitude. Extensive discussions of case S3 are provided in \cite{Nelson2011, Nelson2013a, Nelson2014}. Case S3 uses a dynamic Smagorinsky subgrid-scale model to achieve very low levels of explicit dissipation relative to our other simulations.The dynamic Smagorinsky model was formulated to extrapolate the local unresolved turbulent dissipation using an assumption of scale-invariant behavior in the inertial range of the turbulence cascade \citep{Germano1991}. Details of its implementation in ASH can be found in Appendix A of \cite{Nelson2013a}. Case D3b, which is otherwise identical to case S3 but uses a uniform enhanced eddy viscosity has roughly 50 times more viscous dissipation at mid-convection zone than case S3 \citep{Nelson2013a}. These references also contain additional information on the differential rotation, magnetic activity cycles, and other features of case S3.

\begin{figure}[t]
\begin{center}
  \includegraphics[width=\linewidth]{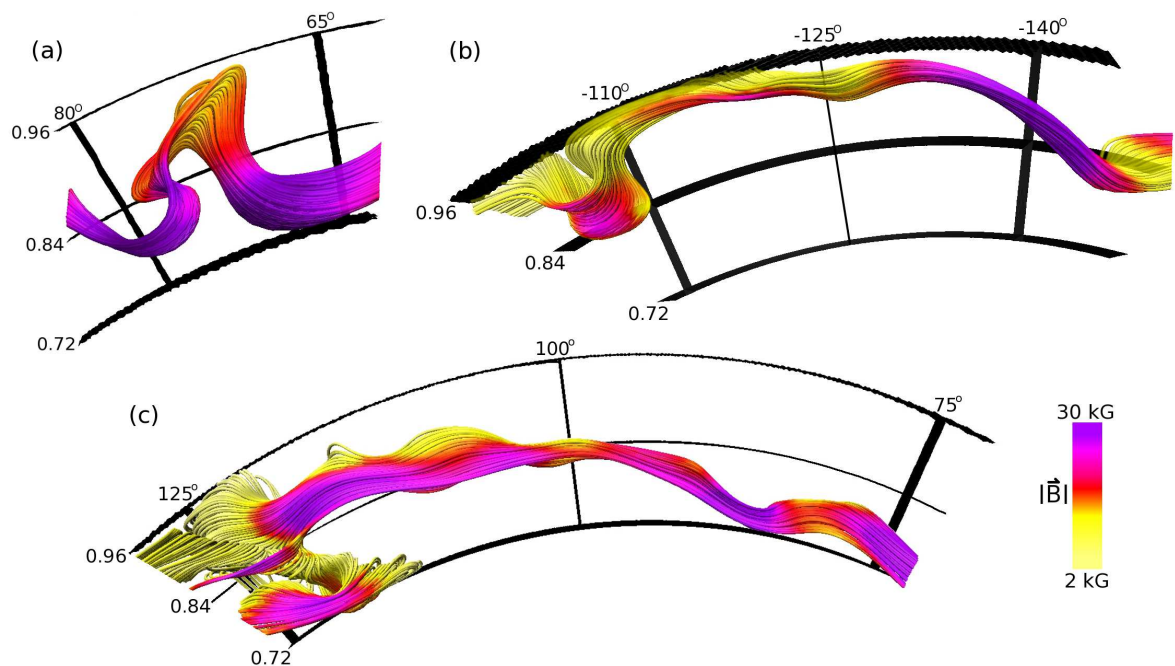}
  \caption{Sample views of three buoyant magnetic loops near the apex of their rises through the convective layer. Loops are shown in 3D volume renderings of magnetic field lines. For all three cases significant flux bundles coherently extend from near the base of the simulated convection zone at $0.72 R_\odot$ to above $0.90 R_\odot$. These loops are shown to highlight the diverse magnetic morphologies found in this simulation.
  \label{fig:Loops}}
  \end{center}
\end{figure}

Case S3 is a turbulent convective dynamo in which strong dynamo action is achieved as convective motions coupled with the bulk rotation of the system lead to strong differential rotation. Figure~\ref{fig:Shells}a displays the convective radial velocities just below mid-convection zone. The convection at low-latitudes displays a strong rotational alignment while the higher latitudes exhibit more isotropic convective patterns. The strong convective shear leads to the generation of magnetic fields with energies at or below the average local kinetic energies and with little large-scale organization. The rotationally-aligned convective cells at low latitudes promote the equator-ward transport of angular momentum, leading to strong differential rotation. Differential rotation, in turn, provides shear which organizes and creates strong toroidal magnetic fields - again at or near equipartition with the local kinetic energy. Figure~\ref{fig:Shells}b shows the existence of strong magnetic wreaths of toroidal magnetic fields. The creation, persistence, and variability of these wreaths have been explored in a variety of settings \citep{Brown2010, Brown2011, Nelson2013a}. These wreaths exhibit strong feedbacks on the convective flows which generate them, leading to regions where the magnetic energy density far exceeds the kinetic energy density. Figure~\ref{fig:Shells}c shows the ratio of magnetic to kinetic energy densities. The cores of the wreaths can achieve ratios above 100. In these areas, Lorentz forces can suppress convective motions. With convective motions suppressed, turbulent mixing in the wreath cores is greatly reduced. We refer to this as reduced resolved turbulent dissipation.

In regions where coherent magnetic structures lead to greatly reduced levels of turbulent dissipation, portions of the wreaths can achieve magnetic fields in the 30 to 50 kG range. Some of these amplified sections can rise through a combination of magnetic buoyancy and favorable advection by convective giant cells and become buoyant magnetic loops. Three such loops are shown in Figure~\ref{fig:Loops}. We have identified nearly 200 of these loops in case S3. Collectively they display statistical properties which mimic observed statistical properties of solar active regions such as Hale's law, Joy's law, the hemispheric helicity rule, and active longitudes \citep{Nelson2014}. These buoyant magnetic loops are flux ropes which are spontaneously generated by convective dynamo action and rise in a self-consistent manner until they are either dissipated or they reach the upper boundary of our simulation, which in case S3 is impenetrable. 

\section{The Formation of Buoyant Loops}

As we have previously demonstrated, these buoyant magnetic loops are coherent magnetic structures that rise through our simulation domain via a combination of magnetic buoyancy and advection by convective giant cells \citep{Nelson2011, Nelson2013a}. As they form in the convection zone, they are never in equilibrium but are constantly experiencing advective forces from the surrounding convection, magnetic and thermal buoyancy, and magnetic tension. In addition, they are continually being re-generated and dissipated by the surrounding flows and our explicit restive diffusion. In many ways, these loops defy the traditional paradigm for buoyant magnetic structures which assumes an initial equilibrium configuration \citep[e.g.,][]{Kersale2007, Vasil2009} or a well-organized and isolated initial magnetic topology \citep[e.g.,][]{Fan2009, Jouve2013}.

Here we will focus specifically on how the segments of the toroidal wreaths that become buoyant magnetic loops are created by dynamo action. To do this we examine the evolution of magnetic energy density in volumes which contain a slice of the loop. We treat this volume in a Lagrangian sense by tracking its motion. We begin by considering the MHD induction equation given by
\begin{equation}
\frac{ \partial \vec{B} }{ \partial t } = \nabla \times \left( \vec{v} \times \vec{B} \right) - \nabla \times \left( \eta \nabla \times \vec{B} \right) .
\label{eq:induction}
\end{equation}
The first term on the right-hand side can be decomposed into three terms which represent advection $-\left(\vec{v} \cdot \nabla \right) \vec{B}$, shearing $\left(\vec{B} \cdot \nabla \right) \vec{v}$, and compression $-\vec{B} \left( \nabla \cdot \vec{v} \right)$. The compression term is further simplified by the anelastic approximation \citep[see][]{Brown2010, Nelson2013a}. The shear term can further be decomposed into contributions from the axisymmetric shear associated with differential rotation and the smaller-scale shearing motions associated with convective flows, and into toroidal and poloidal components. The evolution of the total magnetic energy density $E_M$ is given by taking the dot product of $\vec{B} / 4 \pi $ with Eqn. \ref{eq:induction}, which yields
\begin{equation}
\frac{ \partial E_M }{ \partial t } = \mathcal{M}_\mathrm{PS} + \mathcal{M}_\mathrm{MTS} + \mathcal{M}_\mathrm{FTS} + \mathcal{M}_\mathrm{AD} + \mathcal{M}_\mathrm{RD} + \mathcal{M}_\mathrm{AC} ,
\label{eq:ME Prod}
\end{equation}
where the production terms are 
\begin{equation}
\mathcal{M}_\mathrm{PS} = \frac{ \vec{B}_\mathrm{pol} }{ 4 \pi } \cdot \left[ \left( \vec{B} \cdot \nabla \right) \vec{v} \right] ,
\end{equation}
\begin{equation}
\mathcal{M}_\mathrm{MTS} = \frac{ B_\phi \hat{\phi} }{ 4 \pi } \cdot \left[ \left( \vec{B} \cdot \nabla \right)  \langle \vec{v} \rangle \right] ,
\end{equation}
\begin{equation}
\mathcal{M}_\mathrm{FTS} = \frac{ B_\phi \hat{\phi} }{ 4 \pi } \cdot \left[ \left( \vec{B} \cdot \nabla \right)  \left( \vec{v} - \langle \vec{v} \rangle \right) \right] ,
\end{equation} 
\begin{equation}
\mathcal{M}_\mathrm{AD} = - \frac{ \vec{B} }{ 4 \pi } \cdot \left[ \left( \left[ \vec{v} - \vec{v}_L \right]  \cdot \nabla \right)  \vec{B} \right] ,
\end{equation} 
\begin{equation}
\mathcal{M}_\mathrm{RD} = - \frac{ \vec{B} }{ 4 \pi } \cdot \left[ \nabla \times \left( \eta_t \nabla \times \vec{B} \right) \right] ,
\end{equation} 
\begin{equation}
\mathcal{M}_\mathrm{AC} = \frac{ v_r B^2 }{ 4 \pi } \frac{ \partial \ln{\bar{\rho}} }{ \partial r } .
\end{equation} 
These terms represent the production and dissipation of magnetic energy due to poloidal shear, mean toroidal shear, fluctuating toroidal shear, advection of magnetic energy into or out of the loop slice, resistive diffusion, and anelastic compression. Angle brackets denote longitudinal means. The mean motion of the loop slice is given by $\vec{v}_L$, thus if the entire volume of the loop moved uniformly $\mathcal{M}_\mathrm{AD}$ would go to zero. The spherically-symmetric background density is given by $\bar{\rho}$ and the magnetic resistivity is given by $\eta_t$. Note that $\eta_t$ in case S3 is computed using the dynamic Smagorinsky subgrid-scale model \citep[see][]{Nelson2013a}. We have chosen to divide the shear term into three parts to examine the relative contributions of toroidal shear from the convection and from the differential rotation, as well as the contribution of poloidal shear. In many mean-field models buoyant magnetic flux is assumed to be generated primarily by axisymmetric toroidal shear associated with differential rotation.

These buoyant magnetic loops are most easily identified when they are at or near their maximum radial extent. Perhaps the greatest difficulty in examining the formation of our buoyant magnetic loops is tracking the fluid volumes which become the tops of these loops backward in time. We would like to trace the origin of the magnetic energy which arrives at the top of the domain. This is perhaps the greatest challenge and the greatest insight which we can extract from this simulation, as it self-consistently generates buoyant magnetic loops from convective dynamo action.

To track slices of the loops from times when they can be identified as radially extended coherent magnetic flux ropes to their origins has required the development of a novel tracking algorithm. We start with a slice over $2^\circ$ in longitude at top of the loop at time $t_n$. We chose $t_n$ at a time near where the loop can be identified using magnetic field lines and the top of the loop extends above $0.90 R_\odot$. The cross-section of the loop in radius and latitude $\mathcal{S}_n$ is initially determined using magnetic field lines (similar to what is shown in Figure~\ref{fig:Loops}). We computed the total magnetic energy as $\mathcal{E}_n =  \int_{\mathcal{V}_n} E_M \; dV $, where $E_M$ is the magnetic energy density and $\mathcal{V}_n$ is the volume define by the surface $\mathcal{S}_n$ in radius and latitude and is $2^\circ$ wide in longitude. We compute the volume-averaged magnetic energy generation terms in Equation~\ref{eq:ME Prod} and the volume-averaged motion of the loop $\vec{v}_L^n$. The center of magnetic energy $\vec{C}_n$ in radius, latitude, and longitude is computed as the geometrical center of the loop slice weighted by magnetic energy density. We also compute the magnetic energy generation terms in Equation~\ref{eq:ME Prod} over this volume. Thus we know the change in the total magnetic energy of our volume at the previous time step.

The center of the loop at time $t_{n-1}$ is then computed with a backward Euler time step by $\vec{c}_{n-1} = \vec{C}_n - ( t_n - t_{n-1} ) \vec{v}_L^n$. We use a surface in radius and latitude determined from a new field line tracing, however since we cannot reliably track individual field lines without much finer temporal data, we generally find that this new surface does not enclose the correct total magnetic energy to account for the magnetic energy present at the previous step plus any change due to the generation terms. We either either add or remove a uniform number of grid cells from the edge of the surface until we obtain a volume which contains the same total magnetic energy as was present at the previous step modulo changes from the production terms. This gives us a new loop profile in radius and latitude $\mathcal{S}_{n-1}$. We compute the center of magnetic energy $\vec{C}_{n-1}$ using the integral method and compare it to the center we computed from the backward Euler time stepping. When the difference between the two centers is less than 5\% of the total displacement, we set the loop center to the average of the two and use the radius and latitude profile determined from the modified magnetic field line contour. This process is then repeated to obtain the loop center and profile at time $t_{n-2}$.

\begin{figure}[t]
\begin{center}
  \includegraphics[width=0.48\linewidth]{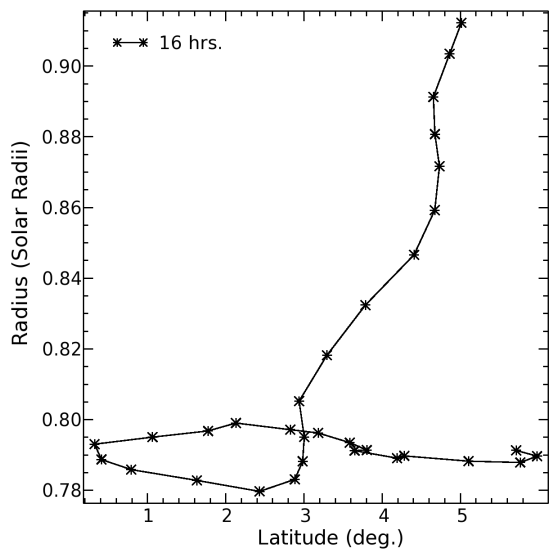}
  \caption{Sample positions of the center of Loop \#1 during its formation and rise through the convection layer taken every 16 hours. The tracked volume begins at $0.791 R_\odot$ and $5.7^\circ$ latitude, and ends 21.6 days later with the uppermost point. The first 14.1 days cover the formation of the loop. The final 7.5 days cover the initial portion of the loop's ascent through the convective layer. The magnetic topologies of the first and last times shown here are displayed in Figure~\ref{fig:Loop1}.
  \label{fig:Tracking}}
  \end{center}
\end{figure}

For situations where the distance between the two centers is larger than 5\% of the total displacement for that time step, we add a third method for determining the loop center and shape. This third method is generally required once the tracked volume enters the lower convection zone where it can become difficult to distinguish from the larger magnetic wreath structure. The algorithm randomly places between 10 and 15 small circular test regions over the loop profile $\mathcal{S}_n$. Each test region is then stepped back in time using the same method used for the entire slice. We compute a new profile in latitude and radius by choosing a profile that contains all of the test regions at time $t_{n-1}$, contains the correct total magnetic energy, and has the minimum distance around its perimeter. We then use this profile to compute the center of magnetic energy using the integral method, which we call $\vec{\mathcal{C}}_{n-1}$. If this method produces a loop center of magnetic energy that lies less than 5\% of the total displacement from both $\vec{c}_{n-1}$ and $\vec{C}_{n-1}$ then we use the average of the three centers and the average of the magnetic energy contour and the test region contour. If $\vec{\mathcal{C}}_{n-1}$ does not lie between $\vec{c}_{n-1}$ and $\vec{C}_{n-1}$ then we deem the method has failed at that step. When the method fails for three successive steps we declare that the tracking algorithm has lost the loop. 

\begin{figure}[t]
\begin{center}
  \includegraphics[width=0.9\linewidth]{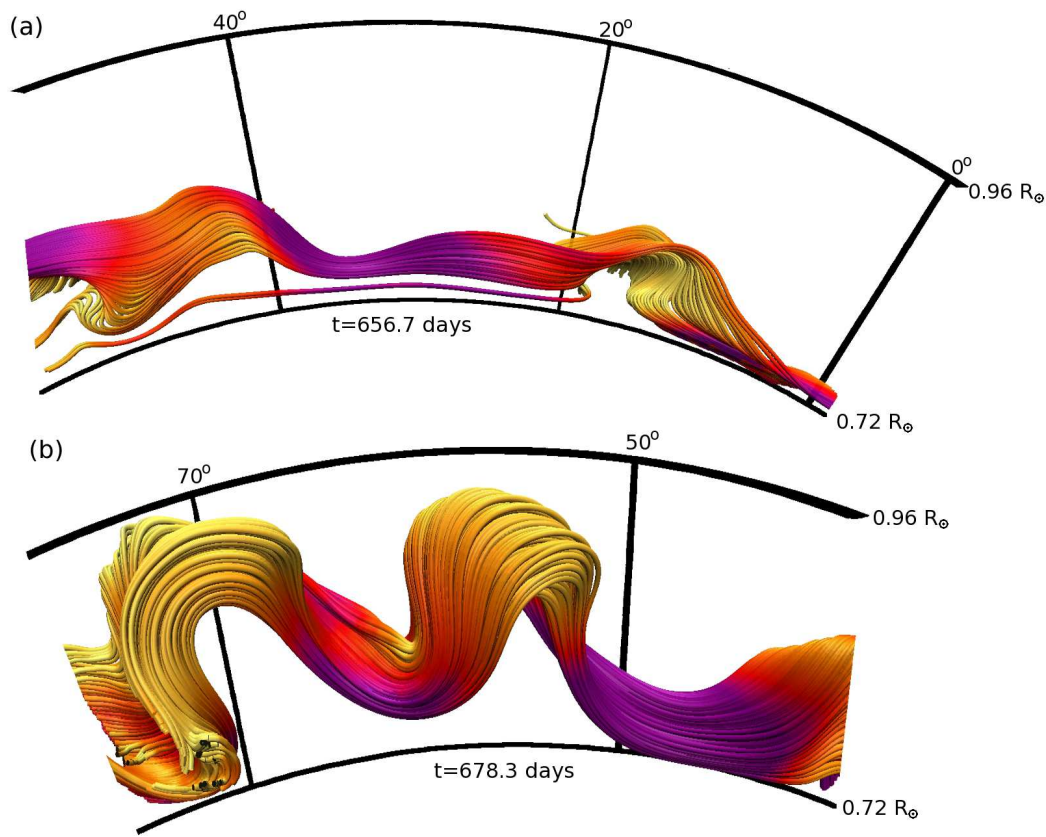}
  \caption{Volume rendering of magnetic field lines passing through the tracking volume for Loop \#1 at the earliest and latest times tracked. Perspective is looking south along the rotation axis with both frames rotated such that radial position increases upwards and longitude increases to the left. Color indicates magnitude of the magnetic field. In both cases the volume being tracked is roughly centered in longitude and outlined by the field lines.
  \label{fig:Loop1}}
  \end{center}
\end{figure}

For this algorithm to work properly, we require the full 3D dataset of the velocity, magnetic field, and dynamic Smagorinsky viscosity at very high temporal cadences. Tests of our tracking algorithm indicate that to successfully track a loop slice over more than 20 days requires time steps of about five hours. There are only six loops for which we have sufficiently fine temporal data over their formation and rise. For one of these six (\#4) the tracking algorithm loses the loop after only 9 days even with our finest temporal cadence. The other five (\#1, \#2, \#3, \#21, \#22) are successfully tracked for at least 20 days. Figure~\ref{fig:Tracking} shows the tracking of Loop \#1 as an example. In the case of Loop \#1, tracking was begun 5 days prior to its maximum radial extent of $0.94 R_\odot$. We do find some sensitivity to the initial times, profiles and total magnetic energies chosen. This sensitivity to initial conditions is minimized when we begin tracking the loop slices before they reach their maximum radial extent. As we can only currently perform this analysis on a small number of loops, we are unable to make general statements about all buoyant magnetic loops in our models.

Figure~\ref{fig:Loop1} shows 3D volume renderings of field line tracings for the first and last times indicated in Figure~\ref{fig:Tracking}. Tracking begins using the large bundle of field lines seen in Figure~\ref{fig:Loop1}(b) at approximately $55^\circ$ longitude. This volume is tracked through the path shown in Figure~\ref{fig:Tracking} backward in time to the magnetic topology shown in Figure~\ref{fig:Loop1}(a). The field that will become the buoyant loop is at that point heavily embedded in the much larger-scale magnetic wreaths which are pervasive at this time and location below about $0.82 R_\odot$.

\begin{figure}[t]
\begin{center}
  \includegraphics[width=0.85\linewidth]{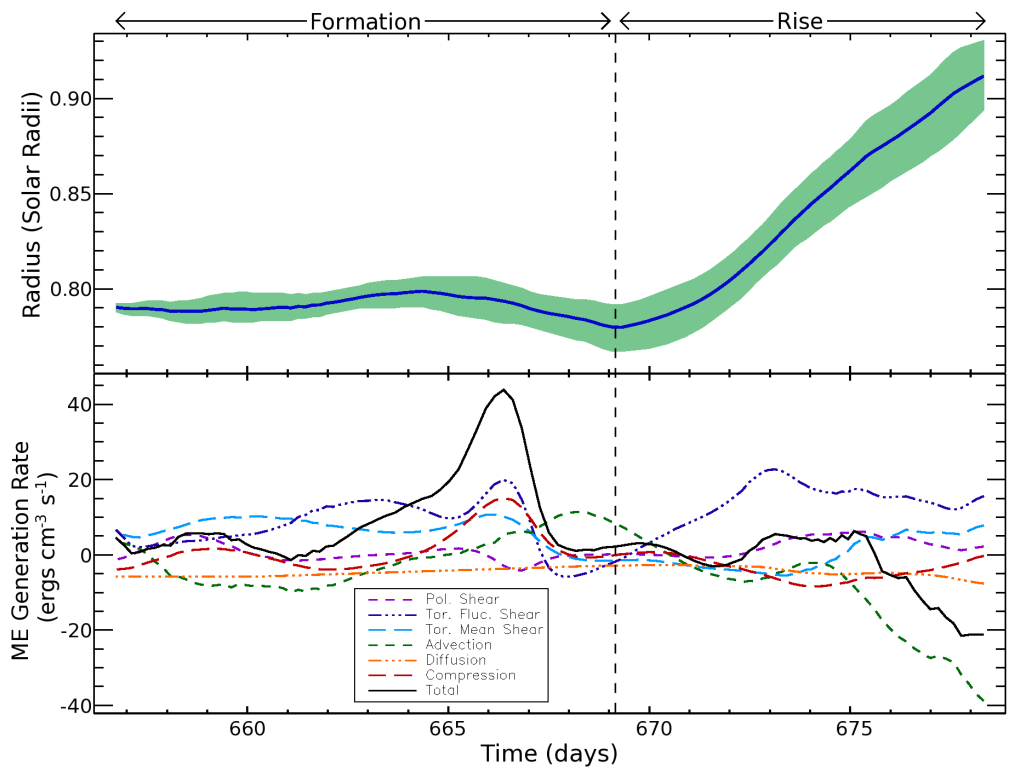}
  \caption{The rise and magnetic energy generation terms as a function of time during the formation and rise of Loop \#1. The top panel shows the radial center (blue line) and radial extent (green shaded region). The rise is deemed to begin once the loop begins monotonically upward motion. The lower panel shows the average magnetic energy generation rate broken down by the terms in Eqn.~\ref{eq:ME Prod}. During the formation stage energy is generated primarily by the fluctuating and mean toroidal shear terms ($\mathcal{M}_\mathrm{FTS}$ and $\mathcal{M}_\mathrm{PTS}$). During the rise the loop's magnetic energy is primarily dissipated by the advection term ($\mathcal{M}_\mathrm{AD}$).
  \label{fig:Generation}}
  \end{center}
\end{figure}

Once we have tracked the loop slices back as far as possible, we then examine the volume-averaged values of the magnetic energy generation terms in Equation~\ref{eq:ME Prod} at each time-step. We identify when the maximum magnetic energy in the tracked volume is achieved and label this $t_\mathrm{max}$. Generally this occurs very close to the beginning of their rise toward the top of our domain. To assess the question of their how these loops are formed, we look at the generation of magnetic energy prior to $t_\mathrm{max}$. For all five of the loops which we tracked successfully at least 13 days of the tracking occurred prior to $t_\mathrm{max}$. By examining the generation of magnetic energy immediately before $t_\mathrm{max}$ we can probe whether the loops are primarily generated through a long, slow build-up of magnetic energy or a rapid amplification on time-scales of about 10 days. For all five loops examined here, we find that between 72\% and 87\% of the magnetic energy present at $t_\mathrm{max}$ was generated during the 13 to 17 days covered by our tracking algorithm. This indicates that these loops are created primarily in a rapid amplification on very short timescales rather than being the product of a slower build-up process.

Figure~\ref{fig:Generation} shows the rise and expansion of a sample magnetic loop as determined by the tracking algorithm, along with the volume-averaged magnetic energy production terms in the loop slice at each time. We divide the time series into the formation phase and the rise phase. The change between the two is deemed to occur when the loop begins monotonically upwards movement. In the upper panel the blue line shows the radial position of the loop center while the green region shows the radial extent of the tracked volume. In the formation phase the radial expansion of the loop is primarily due to increased accumulation of magnetic energy. In the rise phase the expansion is largely the result of the decreasing background pressure at higher radial positions.

A careful examination of the formation phase shows that magnetic energy is primarily generated through the toroidal shear terms $\mathcal{M}_\mathrm{FTS}$ and $\mathcal{M}_\mathrm{MTS}$, with poloidal shear $\mathcal{M}_\mathrm{PS}$ playing a small role through most of the formation phase. All of these terms are highly variable with temporal standard deviations on the order of their temporal means. The formation of Loop \#1 is opposed primarily by restive diffusion, which is small but persistent. Advection also provides a negative net contribution, but it is again highly variable and becomes strongly positive for nearly two days at the end of the formation phase. The compression term predictably follows the radial motion of the loop as it moves slightly upward and then downward during this phase.

The rise phase exhibits some noticeably different behaviors. The fluctuating toroidal shear grows to large positive values as the loop moves radially upward while the mean toroidal shear term drops as the loop moves out of the region of peak differential rotation shear in the bottom of the convection zone before recovering slightly during the later stages of the rise phase. The poloidal shear term increases, again becoming largest in the later stages of the loop's rise.  The compression term becomes more negative as the loop rises. The diffusion term is again far less variable than the other terms but becomes slowly more negative as the average resistivity increases with radial position. The most striking change is the small-scale advection of magnetic energy out of the tracked volume. As the loop rises and leaves the magnetically-dominated region in and around the wreath, it encounters a significant increase in small-scale motions which tend to remove strong fields and replace them with weaker, less-coherent fields from the surrounding fluid. Over the first five days of the rise phase the advective term shows little change, but upon reaching roughly mid-convection zone which marks the rough upper edge of the wreath, the advective term rapidly becomes more negative. By the last time considered here, the advective term has become larger in magnitude than any other term by more than a factor of two.

\begin{figure}[t]
\begin{center}
  \includegraphics[width=0.8\linewidth]{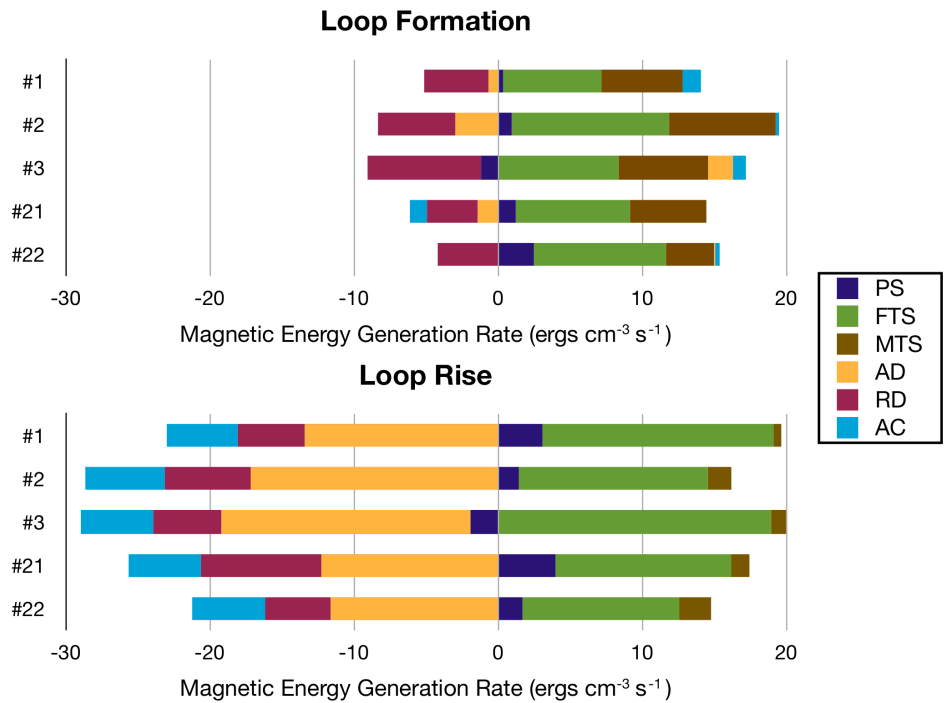}
  \caption{Time- and volume-averaged magnetic energy generation rates for five buoyant magnetic loops (\#1, \#2, \#3, \#21, and \#22) tracked in the same manner as Loop \#1 (see Figs.~\ref{fig:Tracking} and \ref{fig:Generation}). Magnetic energy generation terms follow Eqn.~\ref{eq:ME Prod}. In the formation phase the largest positive contribution comes from small-scale toroidal shearing of poloidal magnetic fields $\mathcal{M}_\mathrm{FTS}$ (green), follow by the differential rotation acting on poloidal magnetic fields $\mathcal{M}_\mathrm{MTS}$ (brown). In the formation stage this is generally opposed by resistive diffusion $\mathcal{M}_\mathrm{RD}$ (magenta). During the rise stage magnetic energy is generated by $\mathcal{M}_\mathrm{FTS}$ (green) at even higher levels than in the formation state, while dissipation is dominated by small-scale advection $\mathcal{M}_\mathrm{AD}$ (yellow) with contributions from $\mathcal{M}_\mathrm{RD}$ (magenta) and anelastic compression $\mathcal{M}_\mathrm{AC}$ (light blue). Poloidal shear $\mathcal{M}_\mathrm{PS}$ (dark purple) plays a small role in both the formation and rise phases.
  \label{fig:5Loops}}
  \end{center}
\end{figure}

We conduct the same analysis of the production and dissipation of magnetic energy on the other four loops for which we have successfully tracked their formation. Figure~\ref{fig:5Loops} shows the time- and volume-averaged values of each generation term for the formation and rise phases of these five loops. While there are significant variations in each loop, some general trends emerge for these five samples. First, in all five cases $\mathcal{M}_\mathrm{FTS}$ is primarily responsible for the generation of magnetic energy in both the formation and rise phases, which $\mathcal{M}_\mathrm{MTS}$ playing a lesser but still significant role in the formation phase. Second, the loops are primarily opposed by resistive diffusion in the formation phase, which is augmented by $\mathcal{M}_\mathrm{AD}$ in the rise phase. With these clear trends, these loops provide strong evidence in support of the essential role of turbulent amplification of magnetic fields in the creation of these buoyant loops and of resolved turbulent dissipation in their eventual dissolution. 

\section{Turbulent Convective Dynamos and Flux Emergence}

In this paper, we have demonstrated that buoyant magnetic loops can be generated in solar-like convective dynamos. The magnetic energy in these loops is primarily generated by the shearing motions of moderate- to small-scale flows with the mean differential rotation shear playing a secondary role. This is in contrast to many dynamo models which assume that buoyant magnetic structures are produced from global-scale toroidal magnetic structures generated by the differential rotation profile. In previous papers we have shown that the axisymmetric shear from the differential rotation plays a key role in the magnetic wreaths and activity cycles in case S3. It is therefore notable that the generation of the buoyant loops analyzed here relies primarily on fluctuating, turbulent shear rather than the axisymmetric differential rotation. 

The ability to capture the spontaneous generation of buoyant magnetic flux ropes in global convective dynamos is an important step forward in understanding how coherent magnetic structures are self-organized by turbulent dynamo action. Additionally, our use of a dynamic Smagorinsky subgrid-scale model has lowered restive diffusion to the point that resolved small-scale advection is responsible for most of the dissipation of these loops. While these loops experience far less turbulent flows than they would in real stellar interiors, it is heartening to see that these loops can survive when turbulent advection becomes the dominant dissipation mechanism.

In a broader context, this work represents an important testing ground for numerical models of magnetic flux ropes generated by turbulent dynamo action. Extensive solar and stellar observations coupled with 3D models are making significant progress in understanding the physical processes whereby stars achieve ordered global behaviors and generate moderate-scale coherent flux ropes from highly chaotic turbulent driving. \\

\section*{Acknowledgements}
We thank Kyle Auguston, Chris Chronopoulos, Yuhong Fan, Nicholas Featherstone, Brandley Hindman, and Joyce Guzik for their suggestions and advice.  This research is partly supported by NASA through
Heliophysics Theory Program grants NNX08AI57G and NNX11AJ36G.
Nelson is supported by a LANL Metropolis Fellowship. Work at LANL was done under the auspices of the National Nuclear Security Administration of the U.S. Department of Engery at Los Alamos National Laboratory under Contract No. DE-AC52-06NA25396. 
Miesch is also supported by NASA SR\&T grant NNH09AK14I.  NCAR is
sponsored by the National Science Foundation.
The simulations were carried out with
NSF TeraGrid and XSEDE support of Ranger at TACC, and Kraken at NICS, and with NASA HECC support of Pleiades.  
Field line tracings and volume renderings shown in Figures~\ref{fig:Loops} and \ref{fig:Loop1} were produced using VAPOR \citep{Clyne2007}.

\bibliographystyle{jphysicsB_mod}
\bibliography{MyCollection}

\end{document}